\documentclass[12pt]{article}  

\usepackage{latexsym}
\usepackage{graphicx}
\usepackage{amsfonts}
\usepackage{amssymb}
\usepackage{amsmath}

\usepackage{multirow}

\def\be{\begin{equation}}
\def\ee{\end{equation}}
\def\ba{\begin{eqnarray}}
\def\ea{\end{eqnarray}}
\def\la{\label}

\renewcommand{\d}{{\rm d}}
\newcommand{\Tr}{\mbox{Tr\,}}

\newcommand{\beq}{\begin{equation}}
\newcommand{\eeq}[1]{\label{#1}\end{equation}}
\newcommand{\bea}{\begin{eqnarray}}
\newcommand{\eea}[1]{\label{#1}\end{eqnarray}}

\newcommand{\e}{{\mathrm e}}
\newcommand{\eg}{{\em e.g.}~}

\newcommand{\eq}{{\,  := \, }}

\newcommand{\GL}{\operatorname{GL}}
\newcommand{\half}{\frac{1}{2}}
\newcommand{\Hdila}{H}

\newcommand{\indJ}{{\text{\tiny J}}}
\newcommand{\indK}{{\text{\tiny K}}}
\newcommand{\indA}{A}
\newcommand{\indB}{B}

\newcommand{\Rset}{\mathbb{R}}
\newcommand{\SL}{\operatorname{SL}}

\newcommand{\SO}{\operatorname{SO}}

\newcommand{\tr}{\operatorname{tr}}

\newcommand{\unit}{\mathbf{1}}
\newcommand{\Vol}{\operatorname{Vol}}
\newcommand{\Zset}{\mathbb{Z}}
\newcommand{\Q}{{\cal Q}}
\newcommand{\Z}{{\cal Z}}
\newcommand{\C}{{\cal C}}

\begin{document}
\begin{flushright}
\hfill{hep-th/0605274} \\
\end{flushright}

\vspace{20mm}

\begin{center}

{\bf Metric and coupling reversal in string theory} \\
\vspace*{0.37truein}
M.J.~Duff\footnote{m.duff@imperial.ac.uk} and J.~Kalkkinen\footnote{j.kalkkinen@imperial.ac.uk} \\
\vspace*{0.15truein}
{\it Blackett Laboratory, Imperial College London} \\
{\it Prince Consort Road, London SW7 2AZ} \\

\end{center}

\baselineskip=10pt
\bigskip

\vspace{20pt}

\abstract{

\noindent Invariance under reversing the sign of the metric
$G_{MN}(x)$ and/or the sign of the string coupling field
${\Hdila}(x)$, where $\langle {\Hdila}(x) \rangle =g_{s}$, leads
to four possible Universes denoted $\unit, {\cal I}, {\cal J},
{\cal K}$ according as $(G,{\Hdila}) \longrightarrow (G,{\Hdila}),
(-G,{\Hdila}), (-G,-{\Hdila}), (G,-{\Hdila})$, respectively.
Universe $\unit$ is described by conventional string/M theory and
contains all M, D, F and NS branes. Universe ${\cal I}$ contains
only D(-1), D3 and D7. Universe ${\cal J}$ contains only D1, D5,
D9 and Type I. Universe ${\cal K}$ contains only F1 and NS5 of IIB
and Heterotic $\SO(32)$.

}

\newpage

\tableofcontents

\newpage


\section{Introduction}
\label{Intro}

\subsection{Signature and coupling reversal}

In a previous paper \cite{Duff2:2006} we identified what class of
field theory is invariant under reversing the sign of the metric
tensor
\ba
G_{MN}(x) \longrightarrow -G_{MN}(x) \la{metricflip}
\ea
induced by a chiral transformation on the curved space gamma matrices
\ba
\Gamma_{M} &\longrightarrow& \Gamma \Gamma_{M} \la{gammaflip} ~,
\ea
where
\ba
\Big\{ \Gamma_M(x) , \Gamma_N(x) \Big\} &=& 2 G_{MN}(x) \unit ~,
\ea
and
$\Gamma$ is the normalised chirality operator
\ba
\Gamma &\equiv & \frac{1}{\sqrt{G}} \frac{1}{D!}\varepsilon^{{M_1}
\cdots M_{D}}\Gamma_{{M_1} \cdots M_{D}} \label{chiral} ~.
\ea
We concluded that theories must be chiral and require signature
$(S,T)$ with $S-T=4k$ in order that the Clifford algebra  be
symmetric. Under (\ref{gammaflip}) the volume element transforms
as
\ba
\sqrt{G} \,\d^Dx & \longrightarrow & (-1)^{D/2}\,\sqrt{G} \,\d^Dx
~,  \la{volflip}
\ea
while the curvature scalar flips sign for all $D$
\ba
R & \longrightarrow & -R ~.
\la{Rflip}
\ea
So for gravitational theories the  requirement of invariance then
selects out the dimensions
\ba
D &=& 4k+2~, \qquad k=0,1,2,3\ldots  \label{a4}
\ea
In D=10, for example, the chiral Type IIB supergravity is invariant while
the non-chiral Type IIA supergravity is not.

In this paper we turn our attention to chiral string theory in
$D=10$. It is now useful to extend the reversal symmetry to
include the dilatonic string coupling field $H(x)$ whose vev is
the string coupling constant\footnote{Note that we are reversing
the field $H(x)$ and not $g_s$, and that the non-zero vev implies
that the symmetry is spontaneously broken \cite{Duff2:2006}.}
\ba
\langle H(x) \rangle  &=& g_{s} ~.
\ea
As for the $B$-field 2-form, we shall see that its reversal is
correlated with reversing $G$ and so we do not treat it as an
independent transformation; the axion $C_0$ transformation is
similarly related to that of $H$, so that the complex parameter
$\tau \equiv C_{0} + i H^{-1}$ transforms homogeneously \footnote{ 
In the quantum theory $\tau$ transforms under $\SL(2,\Zset)$. It may seem unusual to find an action of a discrete 
group $\SL(2,\Zset)$ on the complex
parameter $\tau$ which takes values not restricted to the
upper half of the complex plane. Interestingly enough, such a situation
was also recently encountered in \cite{Kapustin:2006pk}. The context was
somewhat different but also involved orientation reversal.}.

We shall show that this leads
to four possible Universes denoted $\unit, {\cal I}, {\cal J},
{\cal K}$ according as $(G,H) \longrightarrow (G,H), (-G,H),
(-G,-H), (G,-H)$, respectively. Universe $\unit$ is described by
conventional string/M theory and contains all M, D, F and NS
branes. Universe ${\cal I}$ contains only D(-1), D3 and D7.
Universe ${\cal J}$ contains only D1, D5, D9 and Type I. Universe
${\cal K}$ contains only F1 and NS5 of IIB and Heterotic
$\SO(32)$.

\begin{table}
\begin{center}
\renewcommand{\arraystretch}{1.8}
\renewcommand{\tabcolsep}{12pt}
\begin{tabular}{|l|cccc|}
\hline
Universe & $\unit$ & ${\cal I}$ & ${\cal J}$ & ${\cal K}$ \\
\hline
Symmetry &  $(G,H)$  & $(-G,H)$ & $(-G,-H)$ & $(G,-H)$ \\
Branes & All &  D(-1), D3, D7 & D1, D5, D9 & F1, NS5 \\
\hline
\end{tabular}
\caption{Symmetries and branes in the four Universes.}
\end{center}
\end{table}

First of all we show in Section \ref{Supergravity} that at the
level of supergravities, Type IIB is compatible with $\unit, {\cal
I}, {\cal J}, {\cal K}$; Type I is compatible with $\unit, {\cal
J}$; Heterotic $SO(32)$ is compatible with $\unit, {\cal K}$ while
Type IIA and Heterotic $E_{8} \times E_{8}$ are compatible only
with $\unit$. However, we need a finer distinction when we
consider branes which we approach from several different points of
view:
\renewcommand{\arraystretch}{1.2}
\begin{center}
\begin{tabular}{ll}
Section \ref{branes} & {\it Coupling to Dirac-Born-Infeld brane actions} \\
Section \ref{kappa} & {\it $\kappa$-symmetry} \\
Section \ref{solitons} & {\it Brane soliton solutions} \\
Section \ref{quantum} & {\it Invariance properties of string loop and $\alpha'$ corrections, and anomalies} \\
Section \ref{algebras} & {\it Supersymmetry algebras}  \\
\end{tabular}
\end{center}
All these viewpoints will lead to the same $\unit, {\cal I}, {\cal J}, {\cal K}$
classification.

\subsection{Metrics and dilatons}
\label{MetDil}

When we reverse the sign of the metric $G$ and/or dilaton $H$, it
is important to know which metric and which dilaton we have in
mind as there are several different ones that appear in string
theory. For these purposes it is useful to consider the six
strings of table \ref{stringtable} (there are no strings in
Universe ${\cal I}$),
\begin{table}
\begin{center}
\renewcommand{\arraystretch}{1.6}
\renewcommand{\tabcolsep}{4pt}
\begin{tabular}{|l|ccccc|}
\hline
& $\unit$ && ${\cal K}$ && ${\cal J}$ \\
\hline
$N=2$ & {Type IIA} & $\stackrel{T}{\longrightarrow}$ & {Type IIB}$_{\text{F1}}$
& $\stackrel{S}{\longrightarrow}$ & {Type IIB}$_{\text{D1}}$ \\
$N=1$ & ${\text{Het}_{E_8 \times E_8}}$ & $\stackrel{T}{\longrightarrow}$ &
$\text{Het}_{\SO(32)}$  & $\stackrel{S}{\longrightarrow}$ & {Type I} \\
Metric & $G_{A}=G_{E}$ && $G_{F}=G_{O}$ && $G_{D}=G_{I}$ \\
Dilaton & $H_{A}=H_{E}$ && $H_{F}=H_{O}$ && $H_{D}=H_{I}$ \\
\hline
\end{tabular}
\caption{There are three metrics and three dilatons.}
\label{stringtable}
\end{center}
\end{table}
which are, however, related under S-duality as in table
\ref{dualtable}.
\begin{table}
\begin{center}
\renewcommand{\arraystretch}{1.2}
\renewcommand{\tabcolsep}{6pt}
\begin{tabular}{|crc|}
\hline
${\cal K}$ && ${\cal J}$  \\
\hline
~~ $G_{F}$ &=& $H_{D}^{-1}G_{D}$ ~~ \\
$H_{F}$ &=& $H_{D}^{-1}$ \\
\hline
$G_{O}$ &=& $H_{I}^{-1}G_{I}$ \\
$H_{O}$ &=& $H_{I}^{-1}$ \\
\hline
\end{tabular}
\caption{S-dualities between metrics and dilatons of  Universes
${\cal K}$ and ${\cal J}$.} \label{dualtable}
\end{center}
\end{table}

Under T-duality, suppressing the Neveu-Schwarz 2-form and the
graviphoton for simplicity,
\ba
(G_A)_{MN} &=&
\begin{cases}
(G_F)_{MN} & \text{ for } M,N = \mu, \nu \\
(G_F^{-1})_{MN} & \text{ for } M,N = m,n
\end{cases} ~,
\ea
where $\mu,\nu$ correspond to spacetime directions and $m,n$ to
the compact directions in which the T-duality is performed. The
dilaton factors are related by
\be
H_{A}^{-1} = \sqrt{\det (G_F)_{mn}} H_{F}^{-1} \la{Tdual} ~.
\ee
There are similar T-duality formulas relating $G_E$ to $G_O$, and
$H_E$ to $H_O$.

In what follows we shall take our reference metric and dilaton to be those of Type IIB$_{F}$, so $(G,H)$ shall always refer to $(G_{F},H_{F})$.


\section{Supergravity}
\la{Supergravity}

\subsection{Type IIB}
\la{B}

The bosonic Type IIB supergravity action is given by
\cite{Polchinski:1998rr}
\ba
S_{IIB} &=& S_{NS}+S_{R}+S_{CS}
\ea
\ba
S_{NS} &=& \frac{1}{2\kappa_{10}^{2}}\int \d^{10}x \sqrt{G}
{\Hdila}^{-2}
\left (R+4{\Hdila}^{-2}(\partial {\Hdila})^{2}-\frac{1}{12}|H_{3}|^{2} \right) \\
S_{R} &=& -\frac{1}{4\kappa_{10}^{2}}\int \d^{10}x  \sqrt{G} \left
(|F_{1}|^{2}+|\tilde F_{3}|^{2}+\frac{1}{2}|\tilde F_{5}|^{2}
\right ) \\
S_{CS} &=& -\frac{1}{4\kappa_{10}^{2}}\int C_{4}\wedge H_{3}\wedge
F_{3} ~,
\ea
where
\ba
F_{p+1} &=& \d C_{p} \\
\tilde F_{3} &=&F_{3}-C_{0}\wedge H_{3} \\
\tilde F_{5} &=& F_{5}-\frac{1}{2}C_{2}\wedge H_{3}+\frac{1}{2}
B_{2} \wedge F_{3}
\ea
and where we must impose the extra self-duality constraint
\ba
\star \tilde F_{5} &=& \tilde F_{5} ~.
\ea
Since both $S_{NS}$ and $S_{R}$ contains field strengths only of
odd rank, it is invariant under signature reversal of the string
metric
\be G_{MN} \longrightarrow -G_{MN} \la{metricB} ~.
\ee
It is also invariant under
\be {\Hdila} \longrightarrow -{\Hdila} \la{K} ~.
\ee

As explained in \cite{Duff2:2006} the fermionic kinetic terms
\ba
\int \d^{D}x ~ \sqrt{G} ~ \overline{\lambda} \Gamma^M D_M \lambda \quad \text{ and } \quad \int \d^{D}x ~ \sqrt{G} ~
\overline{\psi_M} \Gamma^{MNP} D_N \psi_P
\ea
can be invariant in signature reversal  only if the fermions are
chiral.  This is the case in   both Type IIB and the Type I
supergravities. In particular in Type IIB the  complex gravitino
and the complex dilatino satisfy
\ba
\Gamma \psi_M &=& +\psi_M \\
\Gamma \lambda &=& -\lambda ~.
\ea
This means that their kinetic terms are invariant under signature
reversal.

The interacting  fermionic structure of the theory is captured in
the supersymmetry transformation rules. It is straightforward to
check \cite{Duff2:2006} that they are form invariant under
signature reversal when the bosonic fields are
transformed suitably as well:

The string frame type IIB supersymmetry algebra
\cite{Bergshoeff:2005ac} turns indeed out to remain form invariant in
three different transformations
\ba
(G,{\Hdila}) \longrightarrow (\alpha G, \beta {\Hdila}) ~,  \label{ab0}
\ea
where $\alpha,\beta =\pm 1$, when the gauge fields are transformed
according to
\ba
C_0 & \longrightarrow & \beta ~ C_0   \label{ab1} \\
B_2 & \longrightarrow & \alpha  ~  B_2  \label{ab2} \\
C_2 & \longrightarrow & \alpha\beta  ~ C_2  \label{ab3} \\
C_4 & \longrightarrow & \beta  ~ C_4 ~.  \label{ab4}
\ea
It follows that the standard RR field strengths, the complex
scalar $\tau \equiv C_{0} + i H^{-1}$ and the T-duality covariant matrix $G+B$ then
transform homogeneously.  The
supercovariant field strengths are left invariant as well under
signature reversal.

We conclude that the Type IIB supergravity is invariant under
change of signature in all three sectors
\ba
(G,{\Hdila}) \longrightarrow (-G,{\Hdila}) ~,~ (-G,-{\Hdila}) ~,~
(G,-{\Hdila}) ~.
\ea

\subsection{Type I and Type IIB$_D$}
\la{1}

The Type I supergravity action is given by
\cite{Polchinski:1998rr}
\be
S_{I} = S_{c}+S_{o}
\ee
where
\ba
S_{c} &=& \frac{1}{2\kappa_{10}^{2}}\int \d^{10}x  \sqrt{G} \left(
{\Hdila}^{-2} \left( R+4{\Hdila}^{-2}(\partial {\Hdila})^{2}
\right)
-\frac{1}{12}|\tilde F_{3}|^{2}  \right) \\
S_{o} &=& -\frac{1}{2g_{10}^{2}}\int \d^{10}x  \sqrt{G}
{\Hdila}^{-1} \Tr|F_{2}|^{2} ~,
\ea
and where
\ba
\tilde F_{3} &=& \d
C_{2}-\frac{\kappa_{10}^{2}}{g_{10}^{2}}\omega_{3} ~, \label{F3}
\ea
and $\omega_{3}$ is the Yang-Mills Chren-Simons 3-form
\ba
\omega_{3} &=& \Tr\left(A_{1}\wedge dA_{1}-\frac{2i}{3} A_{1}
\wedge A_{1} \wedge A_{1} \right) ~.
\ea
At first sight this seems not to be invariant under signature flip
of the Type I metric
\be
G_{MN} \longrightarrow -G_{MN} \la{1metric}
\ee
because the Yang-Mills term $S_{o}$ involves field strengths of
even rank. However, in contrast to $S_{c}$, this term is linear in
{\Hdila}, so if we simultaneously perform a coupling flip
\ba
{\Hdila} &\longrightarrow& -{\Hdila} \la{1K}
\ea
then the bosonic part of the Type I supergravity is invariant. The
transformations (\ref{1metric}) and (\ref{1K}) imply that the RR
gauge field $C_2$ is left invariant under signature reversal; it
transforms therefore in the same way as the Chern-Simons term in
(\ref{F3}) leaving $\tilde F_{3}$ invariant.

The simplest way to understand the structure of fermion
interactions in Type I supergravity is to notice that the
closed string sector can be understood as a truncation of the IIB
supergravity theory
\ba
C_{0} = C_{4} = B_{2} &=& 0 ~.
\ea
This means in particular that the Type I string and five-brane are
the Type IIB D-string and D5-brane. The $N=2$ doublet of real
fermions $f^i = \psi^{i}_{M}, \lambda^{i}$ in Type IIB are already
chiral, and in Type I one sets
\ba
(\unit \pm J) f &=& 0 ~.
\ea
The fact that closed string sector fermionic interactions are
invariant follows from the fact that Type IIB supergravity has
signature reversal symmetry.

The open string sector in Type I arises in the Type IIB picture by
adding a suitably symmetrised set of D9-branes. This determines,
in particular, that the gauge sector including the Yang-Mills
fields and the gaugino kinetic term is all multiplied by the same
dilaton factor $H^{-1}$ as what played an important r\^ole in the
bosonic action. As the dilaton factor changes sign in reversal of
signature, we find the opposite chirality condition for the
kinetic term of the gaugino to that of the dilatino.

We conclude that the Type I supergravity is invariant under
change of signature in Universe ${\cal J}$
\ba
(G, H) & \longrightarrow & (-G, -H) ~.
\ea

\subsection{Heterotic $\SO(32)$ and Type IIB$_F$}

The Heterotic $N=1$ supergravity in $D=10$ with gauge group
$\SO(32)$  has the bosonic action
\ba
S_{het} &=& \frac{1}{2\kappa_{10}^{2}}\int \d^{10}x  \sqrt{G}
{\Hdila}^{-2} \left (R+4{\Hdila}^{-2}(\partial
{\Hdila})^{2}-\frac{1}{12}|\tilde H_{3}|^{2}
-\frac{\kappa_{10}^{2}}{g_{10}^2} ~ \Tr|F_{2}|^{2}\right)
\nonumber \\
\ea
where
\ba
\tilde H_{3} &=& \d
B_{2}-\frac{\kappa_{10}^{2}}{g_{10}^{2}}\omega_{3} ~.
\ea

Since the Yang-Mills term is quadratic in {\Hdila}, Heterotic
$\SO(32)$ supergravity requires no flipping the sign of the
hetrotic metric
\be
G_{MN} \longrightarrow G_{MN} \la{Hmetric}
\ee
though it is invariant under coupling flip
\be
{\Hdila} \longrightarrow -{\Hdila} \la{HK} ~.
\ee
The Heterotic $\SO(32)$ supergravity is therefore  invariant under
the symmetry
\ba
(G,H) & \longrightarrow & (G, -H) ~.
\ea

\subsection{Type IIA and  Heterotic $E_8 \times E_8$}
\la{A}

The Type IIA supergravity has the bosonic action
\be
S_{IIA}= S_{NS}+S_{R}+S_{CS}
\ee
where
\ba
S_{NS} &=& \frac{1}{2\kappa_{10}^{2}}\int \d^{10}x  \sqrt{G_{A}}
{\Hdila}^{-2}_A \left (R+4{\Hdila}_A^{-2}(\partial
{\Hdila}_A)^{2}-\frac{1}{12}|H_{3}|^{2}
\right) \\
S_{R} &=& -\frac{1}{4\kappa_{10}^{2}}\int \d^{10}x  \sqrt{G_{A}} \left
(|F_{2}|^{2}+|\tilde F_{4}|^{2} \right ) \\
S_{CS} &=& -\frac{1}{4\kappa_{10}^{2}}\int B_{2}\wedge F_{4}\wedge
F_{4}~,
\ea
and where
\ba
\tilde F_{4} &=& \d C_{3}-C_{1} \wedge H_{3} ~.
\ea
Only even powers of the dilaton $\Hdila_A$ enter the action in
$S_{NS}$, so that the Lagrangean is invariant under
\be
{\Hdila}_A \longrightarrow -{\Hdila}_A \la{KA} ~.
\ee
However, since $S_{R}$ contains RR field strengths of even rank,
it is not invariant under signature reversal.

As the gravitino and the dilatino have components of both
chiralities, the fermionic action will not be invariant in Type
IIA. Apart from the kinetic terms, this can be seen by considering
supersymmetry in opposite signatures. The supersymmetry
transformation rule for the graviton is in string frame
\ba
\delta G_{MN} &=& \overline{\varepsilon} \Gamma_{(M} \Psi_{N)} ~.
\ea
Under the reversal of signature in 10D the graviton $\delta
G_{MN}$ variation maps to
\ba
\overline{\varepsilon} \Gamma_{(M} \Psi_{N)} & \longrightarrow & -
\overline{\varepsilon} \Gamma_{(M} \Gamma\Psi_{N)} ~.
\label{flipGA}
\ea
This is consistent with $\delta G_{MN} \longrightarrow - \delta
G_{MN}$ only when gravitini have  positive chirality in 10D.  The
NS 2-form has to be odd as its supersymmetry transformation reads
\ba
\delta B_{MN} &=& - \frac{\sqrt{2}}{8}    ~ \overline{\varepsilon}
\Gamma_{MN} \lambda - \frac{\sqrt{2}}{4} H_A^{-2} ~
\overline{\varepsilon} \Gamma_{N}\Gamma \psi_{M} \\
&=& - \frac{\sqrt{2}}{8}    ~ \overline{\varepsilon} \Gamma_{MN}
\lambda - \frac{\sqrt{2}}{4} H_A^{-2} ~ \overline{\varepsilon}
\Gamma_{N} \psi_{M} ~.
\ea
For the RR 3-form we get
\ba
 \delta C_{MNP} &=& - \frac{\sqrt{2}}{8}   ~
\overline{\varepsilon} \Gamma_{[MN} \psi_{P]} \\ &=& 0
 ~.
\ea
The Neveu-Schwarz 2-form is naturally odd in signature reversal
but, given the chiral structure of the theory, the 3-form has to
vanish. This can be seen by considering the gravitino
transformation
\ba
\delta \psi_M &=& D_M(\hat\Omega) \varepsilon +
\frac{\sqrt{2}}{288} H_A^{2}~ \Big( {\Gamma_M}^{NKL} - 8
\delta_M^N
\Gamma^{KL} \Big) \Gamma   \varepsilon {H}_{NKL} \\
&& + \frac{\sqrt{2}}{288} \Big( {\Gamma_M}^{NKLP} - 8 \delta_M^N
\Gamma^{KLP} \Big) \varepsilon {F}_{NKLP} ~,
\ea
where the term proportional to ${F}_{NKLP}$ has negative
chirality. The invariant sector of the theory has therefore
\ba
{F}_{NKLP}  &=& 0 ~.
\ea

The same results follow from considering supercovariant quantities
or the fermion interactions in the Type IIA Lagrangean. The
symmetric subsector is given by restricting the Type IIA fields as
follows:
\ba
\Gamma \psi_M &=& +\psi_M \\
\Gamma \lambda &=& -\lambda \\
C_{M} &=& 0 \\
C_{MNP} &=& 0 ~.
\ea
This breaks $N=2$ to a $N=1$ supergravity.

In this subsector we can define the following formal invariance:
\ba
\Vol(M_{10}) & \longrightarrow &  +\Vol(M_{10}) \\
G_{MN} & \longrightarrow & - G_{MN} \\
{{\Hdila}^{-2}_{A}} & \longrightarrow & - {{\Hdila}^{-2}_{A}} \\
H_{MNP} & \longrightarrow & - H_{MNP} ~.
\ea
That the Type IIA is invariant under this operation can be seen
also by T-dualising the signature reversal symmetry of Type IIB
supergravity as will be shown in section \ref{B}. As the dilaton
field becomes imaginary in it, it is not a true symmetry of a
fixed real form of the Type IIA supergravity, however.\footnote{If
one were to adopt a more liberal attitude to the allowed set of
transformations, for example of the kind discussed in section 1.3
of \cite{Duff2:2006}, one may be able to produce a finer
classification of Type IIA branes and interactions beyond those of
Universe ${\unit}$. \label{foot}}

To summarise, we have found that a formally invariant sector of
the Type IIA supergravity which has $N=1$ supersymmetry. As the
action of signature reversal matches the one that can be obtained
from Type IIB by T-dualising, we identify the $N=1$ theory with
the Heterotic supergravity with gauge group $E_{8} \times E_{8}$.

Therefore, neither Type IIA nor  Heterotic $E_8 \times E_8$ are
invariant under change of signature. The reason for this is
fundamentally that the dilaton becomes imaginary in the formal
operation that corresponds to signature reversal in Type IIB.

\subsection{T-duality}
\label{T}

Suppose the background has an isometry along the ninth coordinate
$x^9 \in S^1$. This means that  there is a T-dual configuration in
Type IIA, with\footnote{In this section we distinguish between
Type IIA and Type IIB fields by subscripts $A$ and $B$.}
\ba
\Gamma_9^{\indA} &=& \left(\Gamma_9^{\indB}\right)^{-1} ~. \label{inversio}
\ea
The other gamma matrices $\Gamma_\mu^{\indA}$ coincide with
$\Gamma_\mu^{\indB}$ when the $B_{9\mu}$ components of the NS
2-form are trivial; it turns out that for the purposes of the
following analysis assuming this is not a restriction. The
dilatons are related by
\ba
{\Hdila}_{\indB} &=& k_{9} {\Hdila}_{\indA} ~,
\ea
where $k_9^{+2} = G_{99}$. The chirality operator $\Gamma$ is
invariant under T-duality.

As argued in previous sections, a change of signature in Type IIB
supergravity amounts to
\ba
\Gamma_M^{\indB} & \longrightarrow & +
\Gamma\Gamma_M^{\indB}
 \label{trans1gam} \\
{\Hdila}_{\indB} & \longrightarrow & \pm{\Hdila}_{\indB} ~. \label{trans1dil}
\ea
In what follows we concentrate on the transformation with the upper positive sign above.

On a circle background this defines formally a symmetry of the Type
IIA supergravity as well
\ba
\Gamma_\mu^{\indA} & \longrightarrow & +
\Gamma\Gamma_\mu^{\indA}  \label{trans2gam} \qquad
\mu =
0, \ldots, 8 \\
\Gamma_9^{\indA} & \longrightarrow & -
\Gamma\Gamma_9^{\indA}   \label{trans29} \\
{\Hdila}_{\indA} & \longrightarrow & i
{\Hdila}_{\indA} ~. \label{trans2dil}
\ea
The change of sign in (\ref{trans29}) is a consequence of
(\ref{inversio}). This is not a symmetry of the Type IIA theory
because it involves a complex dilaton field (\ref{trans2dil}).
{See, however, footnote \ref{foot}}.

A further duality along a Killing direction $x^8 \in S^1$ back to
Type IIB changes one of the gamma matrices and the dilaton
\ba
\tilde\Gamma_8^{\indB} &=& \left(\Gamma_8^{\indA}\right)^{-1} = \left(\Gamma_8^{\indB}\right)^{-1} \\
\tilde{\Hdila}_{\indB} &=& \frac{k_9}{k_8}
{\Hdila}_{\indB} ~,
\ea
where $k_8^{-2} = G_{88}$. This induces a symmetry in the new Type
IIB variables
\ba
\tilde\Gamma_\mu^{\indB} & \longrightarrow & +
\Gamma\tilde\Gamma_\mu^{\indB}  \label{trans3gam}
 \qquad \mu = 0, \ldots, 7\\
\tilde\Gamma_i^{\indB} & \longrightarrow & -
\Gamma\tilde\Gamma_i^{\indB}  \label{trans389}
\qquad i=8, 9 \\
\tilde{\Hdila}_{\indB} & \longrightarrow & -
\tilde{\Hdila}_{\indB} ~. \label{trans3dil}
\ea
The action of also this transformation differs from a covariant
operation by a reflection of $(x^8,x^9) \mapsto (-x^8,-x^9)$. Note
that this operation preserves the orientation. We are therefore
back to a symmetry of a Type IIB supergravity, although with the
dilaton changing sign as well.

Quite generally, $2n$ simultaneous  T-dualities take us from the
standard transformation (\ref{trans1gam}) -- (\ref{trans1dil}) to
\ba
\tilde\Gamma_\mu  & \longrightarrow & +
 \Gamma\tilde\Gamma_\mu   \label{trans4gam}
 \qquad \mu = 0, \ldots, 9-2n \\
\tilde\Gamma_i  & \longrightarrow & - \Gamma\tilde\Gamma_i
\label{trans4other}
\qquad i=10-2n, \ldots, 9 \\
\tilde{\Hdila}_{\indB} & \longrightarrow & (-1)^n
\tilde{\Hdila}_{\indB} \label{trans4dil}
\ea
in the new T-dual Type IIB fields.


\section{Coupling to branes}
\la{branes}

\subsection{F1 and NS5 branes}
\la{FNS}

The Nambu-Goto action of a Type IIB or a Heterotic $\SO(32)$
fundamental  string, both with $d=2$, is given by
\ba
S^{F}_{2} &=& -T_{2} \int \d^{2}\xi ~ \sqrt{g} ~,
\ea
where $g_{ij}$ is the induced metric on the brane
\ba
g_{ij} &=& \partial_{i}X^{M}\partial_{j}X^{N}G_{MN}(X) ~,
\label{pbm}
\ea
and in the worldvolume signature $(s,t)$ we define again
\ba
\sqrt{g} &\equiv& \sqrt{(-1)^t \det g_{ij}} ~.
\ea
The Nambu-Goto action of an NS5-brane, with $d=6$,
\ba
S^{F}_{6} &=& -T_{6} \int \d^{6}\xi ~ H^{-2} \sqrt{g} ~,
\ea
depends similarly of only even powers of the dilaton factor $H$.

Due to (\ref{pbm}), a signature reversal in the bulk  pulls back
to a signature reversal on the brane
\ba
g_{ij} &\longrightarrow & - g_{ij}   \label{wvs} ~.
\ea
The volume element on the brane $\sqrt{g}  ~ \d^d \xi$ is built
out of the same pull-back metric, and a choice of reference
orientation on the brane $\epsilon^{i_1 \cdots i_d}$. To
understand its behaviour under signature reversal we express it in
terms of the Clifford algebra valued representation
\ba
\gamma &\eq& \frac{1}{\sqrt{g}} \frac{1}{d!} ~
\partial_{i_1}X^{M_1} \cdots
\partial_{i_d}X^{M_d}~  \epsilon^{i_1 \cdots i_d} ~ \Gamma_{M_1 \cdots
M_d} ~.
\ea
This object squares to $\gamma^2 = \pm \unit$, and has a
fundamental r\^ole in $\kappa$-symmetry in section \ref{kappa}.
For $\gamma$ to have real eigenvalues, we impose a restriction on
the signature of the brane worldvolume
\ba
s-t &=& 4l'
\ea
for some integer $l'$. The fact that in the bulk we have already
required $S-T = 4k'$ guarantees that the  representation of the
Clifford algebra does not change in signature reversal.

Under signature reversal in the bulk, the Clifford algebra valued
orientation transforms as
\ba
\gamma & \longrightarrow & (-1)^{t} \gamma ~,
\ea
and we deduce
\ba
\sqrt{g}  ~ \d^d \xi & \longrightarrow & (-1)^{t} \sqrt{g} ~ \d^d
\xi ~.
\ea
This result is consistent with the na\"\i ve counting as
\ba
(-1)^t &=& (-1)^{d/2}
\ea
in these signatures.

This means that the F1 string and the NS5-brane with Minkowski
signature are never invariant under signature reversal on the
worldvolume, and belong to the Universe ${\cal K}$, where
\ba
(G,H) & \longrightarrow & (G, -H) ~.
\ea

\subsection{Type I and D-branes}
\label{ID}

The coupling of Type IIB supergravity to a D-brane with
worldvolume dimension $d$ is given by the action
\cite{Polchinski:1998rr}
\ba
S^{D}_{d} &=& -T_{d} \int \d^{d}\xi ~ H^{-1} \sqrt{g} ~.
\ea

Using the na\"\i ve rule (\ref{volflip}) for the volume element,
and without requiring that the signature should satisfy any
conditions, this is invariant under changes of spacetime signature
for
\ba
d &=& 4k~, \qquad k=0,1,2,3\ldots \label{4l}
\ea
a multiple of four. In $d=4k+2$ the action can be made invariant
by transforming the dilaton factor $H(x)$ as well. In order to
justify this ad hoc result, we really need the $\kappa$-symmetry
argument of section \ref{kappa}: It will turn out to lead to the
same result.

As the dilaton factor in the Dirac-Born-Infeld action $H$ is a
square root of the dilaton factor $H^2$ that generates the
perturbation theory of the closed oriented Type IIB string, we
have the freedom to choose its transformation properties
independently
\ba
H & \longrightarrow & \pm H\label{dilflip}
\ea
without affecting the supergravity theory in the bulk.

D-brane worldvolume actions are symmetric under the change of
signature provided the sign of $\Hdila$ compensates
for the sign coming from the volume element. We have discussed the
behaviour of the volume element in detail in terms of the bulk
reflection in (\ref{gammaflip}), and in terms of the  brane
reflection in (\ref{braneflip}).

Depending on the sign (\ref{dilflip})  the D-brane theory now
decomposes to two independent sectors:
\begin{center}
\parbox{.8\linewidth}{
\begin{itemize}
\setlength{\topsep}{1.3\topsep}
\setlength{\itemsep}{2\itemsep}
\item[${\cal I}$.  ] For $(G,H)  \longrightarrow  (-G,H)$
we have the D(-1), D3, and D7 branes and the double T-duals of D1, D5, and D9.
\item[${\cal J}$.  ]   For $(G,H)  \longrightarrow  (-G,-H)$ we have the D1, D5, and D9 branes and the double T-duals of D3,
and D7.
\end{itemize}
}
\end{center}
In the former case S-duality is a symmetry of the remaining
theory, whereas in the latter case it is not. S-duality exchanges
Universes ${\cal J}$ and ${\cal K}$.

Note that, in spite of the flip in $\Hdila$, the brane tension always 
remains positive.

The Nambu-Goto action for a Type I string is also of the D1 type
with
\ba
S^{D}_{2} &=& -T_{2} \int \d^{2}\xi ~ H^{-1} \sqrt{g}
\ea
as may be seen from the S-duality rules in section \ref{MetDil}; the Type I string inhabits, therefore, Universe ${\cal J}$.


\section{$\kappa$-symmetry and the brane orientation}
\label{kappa}

There is a fermionic gauge symmetry on the brane,
$\kappa$-symmetry, that combines with the global supersymmetry in
the bulk to induce a global supersymmetry on the brane. Branes in
various signatures are discussed in
\cite{Blencowe:1988sk,Hull:1998fh}.  Kappa-symmetry is chiral, in
a certain sense, as the transformation rules can be written in
terms of a product structure $\hat\gamma$ in the form
\ba
\delta\theta &=& (1+\hat\gamma) \epsilon ~.
\ea
The product structure satisfies $\hat\gamma^2 = \unit$ and $\tr
\hat\gamma = 0$.

For fundamental super $p$-branes such a product structure can be
expressed simply in terms of the pull-back $X^M(\xi)$ and a
worldvolume permutation symbol (orientation)
\cite{Achucarro:1987nc}. It is precisely the Clifford matrix
\ba
\gamma &\eq& \frac{1}{\sqrt{g}} \frac{1}{d!} ~
\partial_{i_1}X^{M_1} \cdots
\partial_{i_d}X^{M_d}~  \epsilon^{i_1 \cdots i_d} ~ \Gamma_{M_1 \cdots
M_d}
\ea
corresponding to world-volume orientation\footnote{There is no
extra complex phase when $p=4l+1, 4l+2$.} for both F1 and NS5.

When the twist field
\ba
{\cal F} &=& F_2^\mathrm{CP} - B_2 \label{twf}
\ea
vanishes, the product structure simplifies to a similar form even
for D-branes \cite{Cederwall:1996pv,Bergshoeff:1996tu}. (A
nontrivial ${\cal F}$ introduces only an even number of gamma
matrices, so our discussion generalises readily to ${\cal F}\neq
0$.) In even dimensions relevant to IIB we have
\ba
\hat\gamma &\eq& \gamma \otimes \tau_d ~.
\ea
Here we have used the notation
\ba
\tau_d &\eq&
\begin{cases}
I & \text{ for } s-t =2,3 \mod 4  \\
J & \text{ for } s-t = 0,1 \mod 4
\end{cases}~.
\ea
The matrix $\hat\gamma$ is automatically traceless, and $\tau_d$
is chosen so that
\ba
\hat\gamma^2 &=& + \unit ~.
\ea
(This is because $\gamma^2 = (-1)^{\half(s-t)(s-t-1)} \unit$ and
$J^2=-I^2=\unit$). The product structure behaves in a change of
signature (\ref{gammaflip}) in even dimensions precisely as
$\sqrt{g} ~ \d^d \xi$ should in  (\ref{volflip}). The reason for
this is the fact $\half d (d-1) = \half d \mod 2$ for even $d$
and, as usual $\Gamma^2=1$.

Given the worldvolume orientation $\epsilon^{i_1 \cdots i_d}$,
$\gamma$ is the volume element of the worldvolume Clifford
algebra. We could not use it directly to implement a worldvolume
change of signature as it does not heed the structure of the
worldvolume embedding in the bulk theory in an appropriate way.
Instead, we should use the {product structure}, and change
signature on the worldvolume by
\ba
\gamma^i \otimes \unit_2 &\longrightarrow & \hat\gamma (\gamma^i
\otimes  \unit_2) ~.
\ea
As it is nilpotent $\hat\gamma^2 =\unit$, it changes the signature
in the Clifford algebra as required. In even dimensions, it
changes itself only by a sign
\ba
\hat\gamma  &\longrightarrow & (-1)^{\half d(d-1)}\hat\gamma ~,
\label{braneflip}
\ea
which reproduces the behaviour of $\sqrt{g} ~ \d^d \xi$ correctly
due to the fact $\half d (d-1) = \half d \mod 2$ when $d$ is even.

The r\^ole played by the product structure in $\kappa$-symmetry is
to carry the information of how the brane is oriented in a
superspace embedding. It therefore incorporates information of the
orientation of the brane together with its coupling to fermions.
This can be made concrete by lifting Type IIB branes into F-theory
using the results of section \ref{F}: the extra $N=2$ structure
becomes geometric, and we see that the product structure
$\hat\gamma$ is precisely the F-theory worldvolume volume element,
as tabulated in table \ref{f-brane}.

\renewcommand{\arraystretch}{1.6}
\renewcommand{\tabcolsep}{4pt}
\begin{table}
\begin{center}
\begin{tabular}{||c|ll|ll|rcl||}
\hline \hline
 \multirow{2}{32pt}{Sector} & \multicolumn{2}{c|}{Type IIB}
 &\multicolumn{2}{c|}{F-theory} & \multicolumn{3}{c||}{Volume} \\
 & Brane & Charge & Brane & Charge & && \\
\hline \multirow{2}{12pt}{${\cal I}$}
& D(-1) & $-$ & ${\cal F}$2 & $\tilde\Z_{12}$ & -${\cal P} \hat\Gamma^{12}$ &$=$& ${\cal P} \otimes I$ \\
& D3 & $Z_{MNP}$ & ${\cal F}$5 & $\tilde\Z_{MNP12}$ & -${\cal P} \hat\Gamma^{MNPQ12}$ &$=$& ${\cal P} \Gamma^{MNPQ} \otimes I$ \\
& D7 & $Z_{M_{1}\cdots M_{7}}$ & ${\cal F}$9 &  $\Z_{M_{1}\cdots
M_{7}12}$& -${\cal P} \hat\Gamma^{M_{1}\cdots
M_{8}12}$ &$=$& ${\cal P} \Gamma^{M_{1}\cdots M_{8}12} \otimes I $ \\
\hline
\multirow{3}{16pt}{${\cal J}$} & D1 & $Z_{M}^{\indJ}$ & ${\cal F}$2 &  $\tilde\Z_{M1}-\Z_{M2}$ & -${\cal P} \hat\Gamma^{MN1}$ &$=$& ${\cal P} \Gamma^{MN}  \otimes J$  \\
& D5 & $Z_{MNPQR}^{\indJ}$ & ${\cal F}$6 &  $\Z_{MNPQR1}$ &
-${\cal P} \hat\Gamma^{MNPQRS1}$ &$=$& ${\cal P} \Gamma^{MNPQRS}
\otimes
J$  \\
& D9 & $Z_{M_1 \cdots M_9}^{\indJ}$ & ${\cal F}$10 &
$\tilde\Z_{M_1 \cdots M_9 1} - \Z_{M_1 \cdots M_9 2}$ & -${\cal P}
\hat\Gamma^{M_1 \cdots M_9 1}$ &$=$& ${\cal P} \Gamma^{ M_1 \cdots
M_9 } \otimes J$
\\ \hline
\multirow{3}{16pt}{${\cal K}$} & F1 & $Z_{M}^{\indK}$ & ${\cal F}$2 &  $\tilde\Z_{M2}   + \Z_{M1}$ & -${\cal P} \hat\Gamma^{MN2}$ &$=$& ${\cal P} \Gamma^{MN} \otimes K$   \\
& NS5 & $Z_{MNPQR}^{\indK}$ & ${\cal F}$6 &  $\Z_{MNPQR2}$ & -${\cal P} \hat\Gamma^{MNPQRS2}$ &$=$& ${\cal P} \Gamma^{MNPQRS} \otimes K$ \\
& F9 & $Z_{M_1 \cdots M_9}^{\indK}$ & ${\cal F}$10 &  $\tilde\Z_{M_1 \cdots M_9 2} + \Z_{M_1 \cdots M_9 1}$ & -${\cal P} \hat\Gamma^{M_1 \cdots M_9 2}$ &$=$& ${\cal P} \Gamma^{M_1 \cdots M_9} \otimes K$ \\
\hline
\multirow{2}{16pt}{All} & KK1 & $P_{M}$ & ${\cal F}$1 &  $\tilde\Z_{M} $ & -${\cal P} \hat\Gamma^{MN}$ &$=$& ${\cal P} \Gamma^{MN} \otimes \unit $   \\
& KK5 & $\tilde{Z}_{MNPQR}$ & ${\cal F}$5 &  $\tilde\Z_{MNPQR}$ & -${\cal P} \hat\Gamma^{MNPQRS}$ &$=$& ${\cal P} \Gamma^{MNPQRS} \otimes \unit$ \\
\hline \hline
\end{tabular}
\end{center}
\caption{Branes and their extensions in Type IIB and F-theory. }
\label{f-brane}
\end{table}

To summarise, we have shown how signature reversal can be
implemented on brane worldvolumes in type IIB theory in an
S-duality covariant way. Taking the $N=2$ structure into account
in this way, leads to the same parity  (\ref{braneflip}) of the
volume element as expected from a complex argument. From the
F-theory point of view the distinction between the three universes
boils down to whether the brane wraps the whole F-theory torus as
in Universe ${\cal I}$; or only the direction along $x^{10}$ as in
Universe ${\cal J}$; or $x^{11}$ as in Universe ${\cal K}$.


\section{Brane solitons}
\la{solitons}

\subsection{Type I and D-branes}
\label{Dbrane}

The solution for extremal D-branes in $D=10$ is \cite{Duff:1994an}
\ba
\d s^{2} &=& \pm \frac{1}{\sqrt{\cal H}} ~ \d x^{2} \pm {\sqrt{\cal H}} ~  \d y^{2} \\
H^{-1} &=& g_{s}^{-1} {\cal H}^{\frac{d-4}{4}}
\ea
where
\ba
C_{d} &=& -({\cal H}^{-1}-1)g_{s}^{-1} \d x^{0}\wedge \d
x^{1}\wedge \ldots \wedge \d x^{d-1} \\
{\cal H} &=& 1+\Big(\frac{y_{d}}{y}\Big)^{8-d} ~.
\ea
The $\pm$ sign in the metric is determined by the boundary
conditions we choose to impose at infinity. Given a harmonic
function ${\cal H} > 0$ that approaches $1$ at infinity, this is
done when we choose the branch of the square root in the metric.

The open string theory depends directly on the dilaton factor $H$
and in particular its sign
\ba
g_s ~ H^{-1} &=&
\begin{cases}
 {\cal H}^{-1}, ~ 1, ~ {\cal H} &
\text{ for D(-1), D3, D7} \\
\pm  (\sqrt{{\cal H}})^{-1}, \pm  \sqrt{{\cal H}} & \text{ for
D1, D5}
\end{cases} ~.
\ea
This means that a change of signature in the metric implies a
change of branch of square root for D1 and D5. The signs of the
dilaton factors of other D-branes D(-1), D3, and D7 do not depend on
a choice of branch and are in fact all fixed given the harmonic
function.

The squareroot ${\cal S} \equiv \pm \sqrt{{\cal H}}$ is a
dynamical degree of freedom that can have positive as well as
negative values. Its value at infinity has to be constant and,
hence, effectively either $1$ or $-1$. This means  that as long as
we do not specify the vacuum expectation value of $\langle {\cal
S} \rangle$, the signature reversal symmetry is not broken. Given
the dependence of the respective dilaton factors on ${\cal S}$, we
have therefore in the unbroken phase the symmetry
\ba
(G, H) & \longrightarrow &
\begin{cases}
(-G, +H)   &
\text{ for D(-1), D3, D7} \\
(-G, -H)  & \text{ for D1, D5}
\end{cases} ~.
\ea
This reproduces the results of section \ref{ID}.

We noted in \cite{Duff2:2006} that metric reversal alone produces
the same effects as sending  $x^M \longrightarrow ix^M$. One may
verify that this does indeed leave the metric, the dilaton and the
$d$-form  invariant in the D(-1), D3, and D7 solutions, where $d=0
\mod 4$.

\subsection{F1 and NS5 branes}
\la{Fbrane}

The solution for extremal fundamental and solitonic branes in
$D=10$ is \cite{Duff:1994an}
\ba
\d s^{2} &=& {\cal H}^{\frac{d-6}{4}}\d x^{2}+{\cal H}^{\frac{d-2}{4}}\d y^{2}\\
H  &=& g_{s}^{}{\cal H}^{\frac{d-4}{4}}
\ea
where
\ba
B_{d} &=& -({\cal H}^{-1}-1)g_{s}^{-1}\d x^{0}\wedge \d
x^{1}\wedge \ldots \wedge \d x^{d-1}
\ea
and where ${\cal H}$ is a harmonic function in the space
transverse to the brane
\ba
{\cal H} &=& 1 + \Big(\frac{y_{d}}{y}\Big)^{8-d} ~.
\ea
Choosing ${\cal H}$ to tend to 1 at infinity rather than -1
involves a choice of boundary condition;  as long as we do not
commit ourselves to a specific signature in $\d x^2$ and $\d y^2$,
this choice does not, as such, break signature invariance yet. The
positive choice, 1, is required to satisfy (\ref{Dila}).

Given the harmonic function ${\cal H} > 0$, the F1 string soliton
solution is
\ba
\d s^{2} &=& {\cal H}^{-1} \d x^{2}+ \d y^{2}\\
H^2  &=&  \frac{ g_{s}^2 }{{\cal H}}
\ea
and the NS5 soliton solution is
\ba
\d s^{2} &=& \d x^{2}+{\cal H} \d y^{2} \\
H^2  &=& g_{s}^2 {\cal H} ~. \label{Dila}
\ea
The closed string effective theory depends of $g_s^2$, and of $H^2
= \e^{2\Phi}$.

The choice of signature in constructing this vacuum  is made when
we choose the signature of the worldvolume metric $\d x^{2}$ and
that of the transverse metric $\d y^{2}$. Changing this choice
affects in no way the sign of the closed string dilaton factor
$H^2$.

This choice of a fundamental string or a Neveu-Schwarz vacuum
breaks the signature reversal symmetry completely. Nevertheless,
as the dilaton factor is given by $H^2 = g_{s}^2 {\cal H} \geq
0$, there is the residual symmetry inherent to any closed string
theory
\ba
(G,H) & \longrightarrow & (G, -H)
\ea
that is the defining characteristic of Universe ${\cal K}$.


\section{Quantum corrections}
\la{quantum}
\subsection{F1 and NS5 branes}

In \cite{Duff2:2006}, we noted that in gravity or supergravity there will be
$L$-loop counterterms of the form
\ba
S_c &\sim& \frac{1}{2\kappa_{D}^{2}}\int \d^D x \sqrt{G} ~
 \kappa_{D}^{2L} ~ R^{\frac{(D-2)L+2}{2}} ~,
\ea
where $R^n$ is symbolic for a scalar contribution of $n$ Riemann tensors
each of dimension 2. These are signature reversal invariant in $D=4k+2$.
String loop corrections in $D=10$ will be of a similar form but with
$\kappa_{10}^{2}$ replaced by $\kappa_{10}^{2}H^{2}$
\ba
S_c & \sim & \frac{1}{2\kappa_{10}^{2}}\int \d^{10} x
\sqrt{G}H^{-2}  ~ \Big(\kappa_{10}^{2}H^{2}\Big)^{L} ~ R^{4L+1} ~,
\ea
and will preserve the signature invariance. However, we noted in
Section \ref{FNS} that the coupling to a fundamental string is not
invariant and so we would expect that the symmetry would not be
respected by $\alpha'$ corrections. This is indeed the case. For
example, at string loop tree level, $L=0$, there are
$\alpha'^{3}R^{4}$ three-loop worldsheet corrections to the Type
IIB string action. Since they involve even powers of $R$, they
violate the symmetry. More generally for F1 and NS5 branes, we
have on dimensional grounds \cite{Duff:1993ye}
\ba
S_{F1} &\sim& \frac{1}{2\kappa_{10}^{2}}\int \d^{10} x \sqrt{G} ~
H^{-2}   ~ \Big(\kappa_{10}^{2}H^{2} \Big)^{L} ~
 T_{2}^{-l}  ~ R^{4L+ l +1} \\
S_{NS5} &\sim& \frac{1}{2\kappa_{10}^{2}}\int \d^{10} x \sqrt{G} ~
H^{-2}   ~ \Big(\kappa_{10}^{2}H^{2} \Big)^{L} ~
\Big(T_{6}^{-1} H^{2} \Big)^{l}  ~ R^{4L+ 3l +1} ~,
\ea
where $l$ is the number of worldvolume loops. So F1 and NS5 branes are
not invariant under signature reversal but only under $(G,H)
\longrightarrow (G,-H)$. Once again we confirm that they belong to Universe $\cal K$.

\subsection{Type I and D-branes}
\label{Icorr}

Under the rules of S-duality of section \ref{MetDil}, however, the
corresponding corrections to the action for a D-Brane or Type I string is obtained by
replacing $T_{d}$ by $T_{d}H^{-1}$:
\ba
S_D &\sim&\frac{1}{2\kappa_{10}^{2}}\int \d^{10} x \sqrt{G} ~
H^{-2}   ~ \Big(\kappa_{10}^{2}H^{2} \Big)^{L} ~
\Big(T_{d}^{-1} H \Big)^{l}  ~ R^{4L+ \frac{dl}{2} +1} ~. 
\ea
Using the rule for the volume element (\ref{volflip}), this is
invariant under $(G,H) \longrightarrow (-G,H)$ only for $d=4k$, but under
$(G,H) \longrightarrow (-G,-H)$ in $d=4k+2$. So we conclude once again that
D(-1), D3 and D7 belong to Universe $\cal I$ while D1,
D5, D9 and the Type I string belong to Universe $\cal J$.

\subsection{Chern-Simons and Green-Schwarz corrections}
\label{GSmechanism}

In the Type I theory and both of the Heterotic theories there is a
Chern-Simons correction to the 3-form Bianchi identities:
\ba
\d \tilde{F}_3 &=&  2 \alpha' X_4  \qquad \text{Type I} \\
\d \tilde{H}_3 &=&  2 \alpha' X_4  \qquad \text{Het}_{\SO(32)} ~,
\ea
where
\ba
2 \alpha' &=& \frac{\kappa_{10}^{2}}{g_{10}^{2}} \\
X_4 & \equiv & \tr R^{2} - \tr F^{2}
\ea
in the notation of \cite{Duff:1990hb}. The anomaly polynomial $X_4$ is
a topological term, and remains
invariant under any signature reversal.  As Type I belongs to the
Universe ${\cal J}$, also the RR 3-form $\tilde{F}_3 $ remains
consistently invariant under metric and coupling reversal, as can be seen
from equation (\ref{ab3}). Similarly, the
3-form $\tilde{H}_3 $ of the Heterotic $SO(32)$ theory remains invariant
under the coupling reversal of Universe ${\cal K}$. In Type IIB
supergravity there is no Chern-Simons correction, and the
Bianchi identity is
\ba
\d H_{3} &=& 0 ~.
\ea
The absence of a Chern-Simons correction is consistent with
S-duality. On the other hand, this is necessary given the transformation rule (\ref{ab3}), and can therefore be seen also as a prediction of
metric reversal invariance in Universe ${\cal I}$.

In Type I and both Heterotic theories there is a Green-Schwarz
anomaly cancellation mechanism that can be expressed as a
modification of the Bianchi identity of the
dual 7-form field strengths  \cite{Duff:1990hb}:
\ba
\d \tilde{F}_7 &=&  - \frac{\beta'}{3} X_8  \label{FF} \\
\d \tilde{H}_7 &=&  - \frac{\beta'}{3} X_8  \label{HH}  ~,
\ea
where
\ba
2 \kappa_{10}^2 &=& \alpha' \beta' (2\pi)^5  \\
X_8 & \equiv & \tr F^{4} - \frac{1}{8} \tr F^{2}
 \tr R^{ 2} + \frac{1}{32} (\tr R^{2} )^{
2} + \frac{1}{8}  \tr R^{4}
\ea
in the notation of \cite{Duff:1990hb}. Once again the anomaly
polynomial $X_{8}$ is signature reversal invariant.
The 7-form field strengths are related to the 3-form field strengths
by
\begin{center}
\begin{tabular}{ll}
$\tilde{F}_7  =   H^{-2} \star \tilde{F}_3$  &\qquad $\text{Type I}$ \\
$\tilde{H}_7   =  \star \tilde{H}_3$  &\qquad
$\text{Het}_{\SO(32)}$.
\end{tabular}
\end{center}
The Hodge star operation changes as
\ba
\star \Sigma_n & \longrightarrow & (-1)^{T+n} \star \Sigma_n ~,
\ea
so operated on odd forms it remains invariant under signature
reversal. The definitions of 7-form fluxes $\tilde{H}_7 $ and
$\tilde{F}_7 $ contain only even powers of the dilaton factor and
remain therefore invariant under coupling reversal. This means that
the  Bianchi identities (\ref{FF}) and (\ref{HH}) are invariant under
the signature reversal of the appropriate universe.  The absence of
Green-Schwarz correction in the Type IIB case
\ba
\d H_{7} &=& 0 ~.
\ea
where
\ba
H_{7}&=& H^{-2} \star H_{3}
\ea
can also be seen as a consequence of metric reversal in Universe ${\cal I}$.


\section{Superalgebras}
\la{algebras}

In this section we consider signature reversal on the level of
superalgebras in 10 dimensions. The only affected algebra
relations are those in which the background metric $\eta_{MN}$ and
the Clifford matrices $\Gamma^M$ enter: these are the relations
that involve the supercharges $Q_{a}$
\ba
\{ Q_{a}, Q_{b} \} &=& (\Gamma^{M}C)_{ab} P_{M} + Z_{ab}
\label{QQgen} \\
 {} [ {L_{M}}^{N}, Q_{a} ] &=& -\frac{1}{4}
{({\Gamma_{M}}^{N}C)_{a}}^{b} Q_{b} ~, \label{LQgen}  \\
  {} [ P_{M}, Q_{a} ] &=& 0 ~,
\ea
where $Z_{ab}$ are central charges. The other commutation
relations define the 10-dimensional  Poincar\'e algebra. This is a
rigid superalgebra, with a constant background metric $\eta_{MN}$
and constant Clifford matrices $\Gamma_M$ that satisfy the
Clifford algebra
\ba
\{ \Gamma_M, ~ \Gamma_N \} &=& + 2 \eta_{MN} ~.
\ea
We have not needed to refer to them anywhere else in the present
work, or in \cite{Duff2:2006}.

There are in fact two possible ways of implementing signature
reversal in the rigid algebra
\ba
\eta_{MN} & \longrightarrow & -\eta_{MN} \label{flat1}\\
\Gamma_M & \longrightarrow & \pm \Gamma \Gamma_M ~. \label{flat2}
\ea
In what follows we shall investigate the symmetry properties of
each algebra for both definitions of signature reversal.

As we are working in dimensions where $\Gamma^{2}=+\unit$, though
$\Gamma_{MN}$ changes sign, after rising an index
${\Gamma_{M}}^{N}$ does not. (This is the reason why we
cannot avoid changing the signature of $\eta_{AB}$ even in the rigid
algebra.) It follows that (\ref{LQgen}) is invariant under change
of signature (\ref{flat1}) -- (\ref{flat2}), when the generators
themselves are left invariant. Postulating that all generators in
the superalgebra are invariant under signature reversal, we find
that the only other algebra relation whose form invariance we need
to check is (\ref{QQgen}).

\subsection{Type I}
\label{Ialgebra}

The Type I superalgebra is common to the Type I and both Heterotic
superstrings. It involves the translation generator $\tilde{P}_M$,
and the self-dual five-form central charge $Z_{MNPQR}^{+}$
\ba
\{ Q_{a}, Q_{b} \} &=& ({\cal P}\Gamma^{M}C)_{ab}  \tilde{P}_{M} +
({\cal P}\Gamma^{MNPQR}C)_{ab} Z_{MNPQR}^{+} ~.  \label{Lp}
\ea
There is no central element for string charges, as the charge
matrix on the left has $136 = 10 + 120$ independent elements. The
chirality projection
\ba
{\cal P}  \equiv \half(\unit - \Gamma)
\ea
is to be thought of as a fixed constant matrix, on equal footing
with the charge conjugation matrix $C$. The minus sign in it is
needed as in our conventions
\ba
\Gamma Z_{5} &=& - \star Z_{5} ~.
\ea

It is straightforward to check that the Type I superalgebra is
invariant under the substitution
\ba
\Gamma^M & \longrightarrow & - \Gamma \Gamma^M \label{upp}
\ea
of (\ref{flat2}). The fundamental reason for this is the presence
of the chirality projection ${\cal P}$.

\subsection{Type IIA}
\label{IIAalgebra}

The Type IIA superalgebra in the 10D Minkowski signature contains the
anticommutation relation \cite{Townsend:1995gp}
\ba
\{ Q_{a}, Q_{b} \} &=& (\Gamma^{M}C)_{ab} P_{M} +  \nonumber \\
&& +  (\Gamma^{M}\Gamma C)_{ab} Z_{M} + (\Gamma^{MNPQR}C)_{ab}
Z_{MNPQR} \\
&& + (\Gamma C)_{ab} Z + (\Gamma^{MN} C)_{ab} Z_{MN} +
(\Gamma^{MNPQ}\Gamma C)_{ab} Z_{MNPQ}   \nonumber ~.
\ea
Choosing the positive sign in (\ref{flat2}), consistent with
(\ref{upp}), the change of signature sends
\ba
P_{M} & \longrightarrow &  - Z_{M} \label{A1} \\
Z_{M} & \longrightarrow &  + P_{M} \\
Z_{MNPQR} & \longrightarrow & \frac{1}{5!} ~
{\epsilon^{M'N'P'Q'R'}}_{MNPQR} ~ Z_{M'N'P'Q'R'} \\
Z, Z_{MN}, Z_{MNPQ} & \longrightarrow & - Z, - Z_{MN}, -Z_{MNPQ} ~. \label{A4}
\ea
The Type IIA algebra is therefore not invariant under change of
signature, as it induces transformations (\ref{A1}) -- (\ref{A4}) on physical charges.

However, the invariant part turns out to be the Type I
algebra (\ref{Lp}) where
\ba
\tilde{P}_{M} &=& \half (P_{M}-Z_{M}) \\
\tilde{Z}_{M} &=& \half (P_{M}+Z_{M}) \equiv 0 ~.
\ea
Choosing the negative sign in (\ref{flat2}), we get the
anti-chiral version of the Type I algebra.

\subsection{Type IIB}
\label{IIBalgebra}

The Type IIB superalgebra contains the anticommutation relation
\cite{Townsend:1995gp}
\ba
\{ Q_{a}^{i}, Q_{b}^{j} \} &=& ({\cal P}\Gamma^{M}C)_{ab} ~ \Big(
\delta^{ij} P_{M} + Z_{M}^{(ij)} \Big) \nonumber  \\
&& + ({\cal P}  \Gamma^{MNPQR}C)_{ab}  ~ \Big(\delta^{ij}
\tilde{Z}_{MNPQR}^{+} + Z_{MNPQR}^{+(ij)} \Big) \\
&& + ({\cal P}\Gamma^{MNP} C)_{ab}  ~ I^{ij} {Z}_{MNP}  ~,  \nonumber
\ea
where $(ij)$ is the traceless symmetric representation of
$\SO(2)$, and ${\cal P}$ is a fixed chirality projection, to be
treated on equal footing with $C$.

When all indices $M,N,P$ are space-like, $Z_{MNP}$ is the D3-brane
charge. The Hodge dual of $Z_{0MN}$ can be written conveniently
using the Clifford representation
\ba
Z_{7} &=& - \Gamma Z_{3} ~,
\ea
and gives the D7-brane charges $Z_{M_{1}\cdots M_{7}}$ when
${M_{1}\cdots M_{7}}$ are space-like. In the notation where
$I=i\sigma_{2}$, $J=\sigma_{1}$ and $K=\sigma_{3}$ we can
decompose the traceless symmetric representation $(ij)$ of
$\SO(2)$ as
\ba
(Z_{n}^{(ij)}) &=& Z_{n}^{\indJ} J +
Z_{n}^{\indK} K
\ea
for $n=1,5$. These central charges are the D1 and the
F1 string charge  for $n=1$, and the D5-brane and the
NS5 brane charge for $n=5$.

The Type IIB algebra is invariant under the action of the automorphism group $\SO(2)$. An example of such automorphisms is
S-duality that acts on supercharges by the transformation $S \in \SO(2)$
\ba
Q^i_a & \longrightarrow & \frac{1}{\sqrt{2}} {( \unit + I
)^{i}}_{j} ~ Q^j ~,
\ea
thus generating an $\Zset_8$ subgroup of $\SO(2)$. S-duality  interchanges the Ramond-Ramond and the Neveu-Schwarz central charges
\ba
(Z_{1}^{\indJ}, Z_{5}^{\indJ}) & \longrightarrow &  (Z_{1}^{\indK}, Z_{5}^{\indK}) \\
(Z_{1}^{\indK}, Z_{5}^{\indK})   & \longrightarrow & (Z_{1}^{\indJ},
Z_{5}^{\indJ}) ~.
\ea

Signature reversal with upper resp.~lower sign in (\ref{flat2})
induces
\ba
\sigma_{\pm}  & : &
\begin{cases}
(Z^{\indK}_{1}, Z^{\indJ}_{1}, Z^{\indK}_{5}, Z^{\indJ}_{5}) &
\longrightarrow  \pm (Z^{\indK}_{1}, Z^{\indJ}_{1}, Z^{\indK}_{5},
Z^{\indJ}_{5}) \\
 Z_{3} & \longrightarrow  \mp  Z_{3} \\
 P & \longrightarrow \pm P
\end{cases} ~.
\ea
We see immediately that, choosing the upper sign,  the
$\sigma_{+}$ transformation is an invariance of the algebra if the
central charges  $Z_{MNP}$ vanish.

Unlike the Type IIA algebra, the Type IIB algebra has  a
nontrivial automorphism group $\SO(2) \subset \GL(2,\Rset)$. As we
shall show in the next section, we can extend signature reversal
$\sigma_{+}$ by including a $\GL(2,\Rset)$ action: the
resulting operations $\hat\sigma_{+}$ preserve different
subsectors of the Type IIB algebra.

The conclusion in the next section will be that  the Type IIB
algebra is not fully invariant under any of the extensions we can
construct. However, the  Type IIB algebras of a given Universe
${\cal I}$, ${\cal J}$, or ${\cal K}$ are invariant under at least
one such extension each.

\subsection{Subsectors of Type IIB}
\label{Bsectors}

One may redefine the supercharges
\ba
Q_{a}^{i} & \longrightarrow & {g^{i}}_{j} ~ Q_{a}^{j}
\ea
by acting on their $N=2$ structure with an element $g \in
\GL(2,\Rset)$. Only the  subgroup $\SO(2) \subset \GL(2,\Rset)$
of these transformations qualifies as actual automorphisms of the
superalgebra. For instance, only $g \in \mathrm{O}(2,\Rset)$
preserve the translation generator $P_{M}$ as the condition for
this is $g^{T}g = \unit$.

We are looking for an extension of $\sigma_{+}$ by combining  with
it  an action of $g \in \GL(2,\Rset)$, such that the resulting
transformation $\hat\sigma_{+}$ leaves at least a part of the Type
IIB algebra form invariant.

Choosing an automorphism  $g \in \SO(2)$ would not lead to a  new
operation $\hat\sigma_{+}$; on the other hand any element  $g \in
\GL(2,\Rset)$ must be orthogonal $g^{T}g = \unit$. This leaves us
precisely two candidates for extensions, namely using $g=J$ or
$g=K$. The actions of these elements (including $I$ for
completeness) on the algebra are
\ba
\nonumber \\
(-)^{F_{L}}  & : &
\begin{cases}
Q_a^{i}  & \longrightarrow  {K^{i}}_{j}Q_{a}^{j} \\
(Z^{\indJ}_{1}, Z_{3}, Z^{\indJ}_{5}) & \longrightarrow  -
(Z^{\indJ}_{1}, Z_{3}, Z^{\indJ}_{5})
\end{cases} \\
\nonumber \\
\Omega  & : &
\begin{cases}
Q_a^{i}  & \longrightarrow  {J^{i}}_{j}Q_{a}^{j} \\
(Z^{\indK}_{1}, Z_{3}, Z^{\indK}_{5}) & \longrightarrow  -
(Z^{\indK}_{1}, Z_{3}, Z^{\indK}_{5})
\end{cases}  \\
\nonumber \\
\varpi  & : &
\begin{cases}
Q_a^{i}  & \longrightarrow {I^{i}}_{j}Q_{a}^{j} \\
(Z^{\indK}_{1}, Z^{\indJ}_{1}, Z^{\indK}_{5}, Z^{\indJ}_{5}) &
\longrightarrow -(Z^{\indK}_{1}, Z^{\indJ}_{1}, Z^{\indK}_{5},
Z^{\indJ}_{5})
\end{cases} ~,
\nonumber \\
\ea
where $(-)^{F_{L}}$ arises in string theory as the left-handed
worldsheet fermion number  modulo two, and $\Omega$ as the
worldsheet parity transformation, see \eg \cite{Bergshoeff:2000qu}.

Central charges that appear with $I$, $J$, or $K$ in the
superalgebra can now be projected out by combining $\sigma_{+}$ with
either $\unit$, $\Omega$, or $(-)^{F_{L}}$. We denote these
extensions of signature reversal by
\ba
\hat\sigma_{I} &=& \sigma_{+} \circ \unit \\
\hat\sigma_{J} &=& \sigma_{+} \circ \Omega  \\
\hat\sigma_{K} &=& \sigma_{+} \circ (-)^{F_{L}} ~.
\ea
The corresponding projections onto invariant subsectors are ${\cal
P}_{I}$, ${\cal P}_{J}$, and ${\cal P}_{K}$.

The projection procedure can be repeated. If we impose invariance
under all three operations  $\hat\sigma_{I}$, $\hat\sigma_{J}$,
and $\hat\sigma_{K}$ simultaneously, we get the projection ${\cal
P}_{I} \circ {\cal P}_{J} \circ {\cal P}_{K}={\cal P}_{I}$, and are
left with the purely gravitational Type I sector in the algebra
\ba
\{ Q_{a}^{i}, Q_{b}^{j} \} &=&  \Big( ({\cal P}\Gamma^{M}C)_{ab} ~
P_{M} + ({\cal P}  \Gamma^{MNPQR}C)_{ab}  ~ \tilde{Z}_{MNPQR}^{+}  \Big)
\delta^{ij}  ~.
\ea
This is clearly the minimal invariant subsector. The five-form
charge here is associated  with a Kaluza-Klein monopole, rather
than a five-brane \cite{Townsend:1995gp}.

By imposing invariance only with respect to a pair of operations,
we may project the algebra  onto sectors that include certain
subsets of brane charges. This leads to three subsectors of the
theory that are separately invariant with respect to two different
extensions of signature reversal:
\begin{itemize}

\item[${\cal I}$.]
The sector invariant under both  $\hat\sigma_{J}$,
and $\hat\sigma_{K}$ has the superalgebra
\ba
\{ Q_{a}^{i}, Q_{b}^{j} \} &=& ({\cal P}\Gamma^{M}C)_{ab} ~
\delta^{ij} P_{M} + ({\cal P}\Gamma^{MNP} C)_{ab}  ~ I^{ij}
{Z}_{MNP}
\nonumber  \\
&& + ({\cal P}  \Gamma^{MNPQR}C)_{ab}  ~ \delta^{ij}
\tilde{Z}_{MNPQR}^{+}  ~.
\ea
This sector involves only Ramond-Ramond D3 and D7 brane charge,
apart from the gravitational charges  $P_{M}$ and $\tilde{Z}_{MNPQR}^{+}$.
This sector is invariant under $\varpi$.

\item[${\cal J}$.]
The sector invariant under both the pure signature reversal
$\sigma_{+}$ and   its extension to $\hat\sigma_{J}$ has the
superalgebra
\ba
\{ Q_{a}^{i}, Q_{b}^{j} \} &=& ({\cal P}\Gamma^{M}C)_{ab} ~
\Big(\delta^{ij} P_{M} + Z_{M}^{\indJ} J^{ij} \Big)
\nonumber  \\
&& + ({\cal P}  \Gamma^{MNPQR}C)_{ab}  ~ \Big(\delta^{ij}
\tilde{Z}_{MNPQR}^{+} + Z_{MNPQR}^{\indJ} J^{ij}
\Big) ~. \label{suJ}
\ea
The charges $Z^{\indJ}_{1}$ are the D-string charge and
$Z^{\indJ}_{5}$ the D5-brane charge. The  two projections onto
states that are invariant under both $\sigma_{+}$ and
$\hat\sigma_{J}$ is actually the same as the projection onto
states that are invariant under $\Omega$.

The brane charges arising in this sector survive the $\Omega$
orientifold projection that projects Type IIB string theory to
Type I $\simeq$ Type IIB$/\Omega$ string theory. We can identify
this sector with the Type I supergravity, even if the algebra in
(\ref{suJ}) has more central elements than
the actual Type I algebra.

\item[${\cal K}$.]
The sector invariant under both the pure signature reversal
$\sigma_{+}$ and   its extension to $\hat\sigma_{K}$ has the
superalgebra
\ba
\{ Q_{a}^{i}, Q_{b}^{j} \} &=& ({\cal P}\Gamma^{M}C)_{ab} ~
\Big(\delta^{ij} P_{M} + Z_{M}^{\indK} K^{ij} \Big)
\nonumber  \\
&& + ({\cal P}  \Gamma^{MNPQR}C)_{ab}  ~ \Big(\delta^{ij}
\tilde{Z}_{MNPQR}^{+} + Z_{MNPQR}^{\indK} K^{ij}
\Big) ~.
\ea
This case is similar to ${\cal J}$, except that it is the $K$-component of the  symmetric central charges that survives. The surviving
central charges  $Z^{\indK}_{1}$ are the  fundamental string charge,
and $Z^{\indK}_{5}$ the Neveu-Schwarz 5-brane charge. This sector is
invariant under $(-)^{F_L}$.

\end{itemize}

\subsection{F-theory formulation}
\label{F}

Let us define the 12-dimensional gamma matrices
\ba
\hat\Gamma^{{M}} &=& \Gamma^{M} \otimes \unit \\
\hat\Gamma^{10} &=& \Gamma \otimes J \\
\hat\Gamma^{11} &=& \Gamma \otimes K ~,
\ea
where $M=0, \ldots, 9$. This Clifford algebra has signature
$(11,1)$ or $(3,9)$, the charge conjugation matrix   $\C = C
\otimes I$, and the chirality operator $\hat\Gamma = -\Gamma
\otimes I$. The 12D algebra is not signature
reversal invariant, as it has $S-T = 2 \mod 4$. The most general
ansatz for an anticommutation relation for a 12D
supercharge $\Q_{a}$ is
\ba
\{ \Q_{a}, \Q_{b} \} &=&
(\hat\Gamma^{\hat{M}} \hat\Gamma \C)_{ab} ~ \tilde\Z_{\hat{M}} +
(\hat\Gamma^{\hat{M}\hat{N}} \C)_{ab} ~ \Z_{\hat{M}\hat{N}} \nonumber \\
&&  +
(\hat\Gamma^{\hat{M}\hat{N}} \hat\Gamma \C)_{ab} ~ \tilde\Z_{\hat{M}\hat{N}}
+
(\hat\Gamma^{\hat{M}\hat{N}\hat{P}} \C)_{ab} ~ \Z_{\hat{M}\hat{N}\hat{P}} \label{Fsu} \\
&& +
(\hat\Gamma^{\hat{M}\hat{N}\hat{P}\hat{Q}\hat{R}} \hat\Gamma \C)_{ab} {\tilde\Z}_{\hat{M}\hat{N}\hat{P}\hat{Q}\hat{R}}
+
(\hat\Gamma^{\hat{M}\hat{N}\hat{P}\hat{Q}\hat{R}\hat{S}} \C)_{ab} {\Z}_{\hat{M}\hat{N}\hat{P}\hat{Q}\hat{R}\hat{S}} ~,  \nonumber
\ea
where $\hat{M}=0, \ldots, 11$ is the 12D index. As $64 \times 64$
matrices we have 2080 free components on both sides.

This structure does not extend to a covariant 12D Poincar\'e
superalgebra \cite{vanHolten:1982mx}. In
signature  $(10,2)$ there is a superalgebra, however, originating
from the supergroup $\mathrm{OSp}(1|32)$, but it has no
translation generators $P_{M}$. In what follows we shall make use
of (\ref{Fsu}) purely on a formal level, and no claim of its 
extending to a superalgebra is made. There is some evidence that
this structure is not entirely accidental, however.  First of all,
F-theory does have the full 32 supersymmetries, and involves a
fundamental 3-brane in an $(11,1)$ background
\cite{Jatkar:1996zk}. Furthermore, as we have seen in section
\ref{kappa}, precisely this matrix structure arises in
$\kappa$-symmetry.

The algebra (\ref{Fsu}) should be understood as convenient  12D
notation for the following 10D central charges:
\ba
P_{M} &=&  \tilde\Z_{M} + 6 ~ \Z_{M 1 2} \\
Z^{(ij)}_{M} &=&  2 ~ \Big( (\tilde\Z_{M\alpha} \unit + \Z_{M\alpha} I  ) \cdot \tau^{\alpha}\Big)^{ij} \\
Z_{MNP} &=& \Z_{MNP} - 20 ~ \tilde\Z_{MNP 1 2}  \\
\tilde{Z}_{MNPQR} &=& \tilde\Z_{MNPQR}  \\
Z^{(ij)}_{MNPQR} &=&  6 ~ \Z_{MNPQR\alpha} (  \tau^{\alpha} I)^{ij} \\
Z_{M_1 \cdots M_7} &=& \Z_{M_1 \cdots M_7} - 72 ~ \tilde\Z_{M_1 \cdots M_7 1 2} \label{Z7} \\
Z^{(ij)}_{M_1 \cdots M_9} &=&  10 ~ \Big( (\tilde\Z_{M_1 \cdots
M_9 \alpha} \unit + \Z_{M_1 \cdots M_9\alpha} I  ) \cdot
\tau^{\alpha}\Big)^{ij} \label{Z9}
\ea
where $M,N,P,Q,R$ are 10D indices, $\alpha,\beta = 1,2$, and
$\tau^{1}=J,  \tau^{2}=K$, and we set other central charges to
trivial values. With these restrictions in effect, the negative
chirality part of (\ref{Fsu}) is the Type IIB superalgebra
\ba
\{ \Q_{a}, \Q_{b} \} &=&  \Big( {\cal P}\Gamma^{M}C  \otimes \unit
~ P_{M} + {\cal P}\Gamma^{M}C  \otimes Z_{M} \Big)_{ab} ~
\nonumber
\\ && + \Big({\cal P}\Gamma^{MNPQR}C \otimes (\tilde{Z}_{MNPQR} ~
\unit +
Z_{MNPQR})\Big)_{ab} \\
&& + ({\cal P} \Gamma^{MNP} C \otimes I )_{ab}  ~ {Z}_{MNP}
\nonumber ~.
\ea

In $D$ dimensions an $n$-form central charge $Z_{M_1 \cdots M_n}$
is related to a $D-n$-form charge $Z_{M_1 \cdots M_{D-n}}$ by 
Hodge duality
\ba
\tilde{Z}_{n} \Gamma  & \equiv & Z_{D-n} ~.
\ea
This provides a relationship between $\tilde{Z}_1, Z_{3}$ and
$Z_9,\tilde{Z}_7$ in 10D, respectively, and $\tilde{\Z}_5, \Z_{2},
\tilde\Z_2$ and $\Z_{7},\tilde{\Z}_{10},\Z_{10}$ in 12D. We have
made use of this structure in identifying the central charges
(\ref{Z7})--(\ref{Z9}).

Implementing the 10D signature reversal in this algebra keeps
central charges $Z_n$ invariant when $n=4k+2$ and reverses the
central charge when $n=4k$. The action of $\mathrm{O}(2)$ can now be understood in terms of reflections in
the two new directions $x^{10}$ and $x^{11}$, so that the three
involutions $(-)^{F_L}$, $\Omega$, and $\varpi$ can be lifted to
geometric actions on this 12D algebra. The sector ${\cal I}$, for
instance, will not involve all of the 12D charges but components
not involving indices in both of the $x^{10}$ and $x^{11}$
directions get projected out
\ba
\Z_{MNP} &=& 0 \\
\Z_{M_1 \cdots M_7} &=& 0 ~;
\ea
only  $\tilde\Z_{MNP 1 2}$ and $\tilde\Z_{M_1 \cdots M_7 1 2}$
survive the projection. Similarly in the sectors ${\cal J}$ and
${\cal K}$, central charges corresponding to branes wrapping
around the direction $x^{10}$ and $x^{11}$ will be projected out.


\section{U-duality}
\label{U}

We have seen that under $(G,H) \longrightarrow (-G,H), (-G,-H),
(G,-H)$,  Type IIB breaks up in three different Universes ${\cal
I}$, ${\cal J}$, ${\cal K}$. T-duality interchanges ${\cal I}$ and
${\cal J}$, S-duality interchanges ${\cal J}$ and  ${\cal K}$.

In the usual Universe $\unit$, for  which $(G,H)\longrightarrow
(G,H)$, compactification to $D=4$ on $T^{6}$ results in a
U-duality group \cite{Cremmer:1979up,Duff:1990hn,Hull:1994ys}
$E_{7(7)}(\Zset)$ under which the  charges transfrom as a
$\mathbf{56}$. Decomposing under the S and T-duality groups,
\be
E_{7(7)}(\Zset) \longrightarrow \SL({2},\Zset) \times
\SO(6,6;\Zset)
\ee
yields
\be
\mathbf{56} \longrightarrow (\mathbf{2},\mathbf{12}) \oplus
(\mathbf{1},\mathbf{32})
\ee
where the $(\mathbf{2},\mathbf{12})$ refers to NS charges and the $(\mathbf{1},\mathbf{32})$ to RR. These
charges have an interpretation in terms of branes wrapped around
$T^{6}$ \cite{Hull:1994ys}. The $(\mathbf{2},\mathbf{12})$ come from the F1 and NS5 while the
$(\mathbf{1},\mathbf{32})$ come from the D-branes. In order to see how these are divided
between the different branes, we decompose further 
\ba
\SO(6,6;\Zset) &\longrightarrow & SL(6,\Zset) \times \Zset_{2}
\ea
under which
\ba
\mathbf{32}  &\longrightarrow &  \mathbf{1}^{+3} \oplus
\mathbf{1}^{-3}  \oplus
\mathbf{15}^{+1} \oplus  \mathbf{15}^{-1} \\
 \mathbf{32}  &\longrightarrow &
\mathbf{6}^{+2} \oplus  \mathbf{6}^{-2}  \oplus \mathbf{20}^{0}
\ea
for IIA and IIB, respectively. Here $ \mathbf{1}^{\pm 3}$ refers to
D0 and D6, $\mathbf{15}^{\pm 1}$ to D2 and D4, $\mathbf{6}^{\pm
2}$ to D1 and D5 and $\mathbf{20}^{0}$ to D3.

This gives rise to the U-duality groups for the various sectors shown in the table:
\renewcommand{\arraystretch}{1.6}
\renewcommand{\tabcolsep}{12pt}
\begin{center}
\begin{tabular}{|c|ccc|}
\hline $\unit$ & ${\cal I}$ & ${\cal J}$ & ${\cal K}$  \\
\hline $E_{7(7)}(\Zset)$ & $SL(6,\Zset)$ & $SL(6,\Zset) $  &
$\SL(2,\Zset) \times
\SO(6,6; \Zset)$ \\
$\mathbf{56}$ & $\mathbf{20}$ &  $\mathbf{6}  \oplus \mathbf{6} $  & $(\mathbf{2},\mathbf{12})$  \\
\hline
\end{tabular}
\end{center}


\section{Yang-Mills: bulk versus brane}
\label{YM}

The existence of  signature reversal invariant Yang-Mills theories in
signature $(3,1)$ is
remarkable given the negative conclusions we reached in
\cite{Duff2:2006}. There, we recall, the kinetic term in pure
Yang-Mills
\ba
S_{YM} &=& \frac{1}{4 g_D^2} \int \d^D x ~ \sqrt{G} ~ \Tr
|F_{2}|^{2}
\ea
contains two contractions with the background metric. Invariance
then requires that the volume form should not change sign under
signature reversal. Consequently pure Yang-Mills theory is invariant
only in dimensions $D=4k$. If the theory is coupled to fermions,
and we require $S-T = 4k'$, this leads to $D=4k'+2T$ so that there
would have to be an even number of time-like dimensions. In
particular, $D=10$ super Yang-Mills is ruled out. However, the
Yang-Mills theory that appears in AdS/CFT is the one appropriate
to Dirac-Born-Infeld D3-brane action
\ba
S^{D}_{4} &=& -T_{4} \int \d^{4}\xi ~ H^{-1} \sqrt{ \det
(g_{ij}+{\cal F}_{ij})} ~,
\ea
where the pull-back metric is defined in (\ref{pbm}) and the twist
field on the brane ${\cal F}_{ij}$ in (\ref{twf}). As we have seen
in section \ref{ID}, this is invariant in $(3,1)$, because the
relevant gamma matrices are those of $D=10$ rather than $d=4$.

There are two other ways in which Yang-Mills in $(3,1)$ is
allowed: first, as shown in  \cite{Duff2:2006}, although forbidden
in $D=4k+2$, Maxwell and Yang-Mills terms can arise after
compactification to lower dimensions. Secondly, their absence in
$D=4k+2$ applies in pure Yang-Mills theory only. In the Yang-Mills
sector of Type I supergravity the kinetic term is multiplied by a
dilaton factor ${\Hdila}(x)$ that may also change sign to
compensate for the change in sign of the volume element
\cite{Duff2:2006}. It would be worthwhile to investigate the
residual effects of $D=10$ signature reversal in 4-dimensional
string and M-theory phenomenological models.


\section{Conclusions}
\label{Conclusions}

We have seen that combined effects of  metric and coupling
reversal give rise to four different universes. Curiously, only
Universe $\unit$ is M-friendly in the sense of admitting M2 and
M5-branes, and their descendants in Type IIA and Heterotic $E_8
\times E_8$. On the other hand, Universe ${\cal I}$ has no strings
at all, but only D(-1), D3, and D7-branes. It is worth pursuing
what such a universe would be like: for example, we still have
AdS/CFT on AdS$_5 \times S^5$, but not on  AdS$_4 \times S^7$ or
AdS$_7 \times S^4$. It could be argued, of course, that there is
an advantage in singling out Yang-Mills theory in $(3,1)$ rather
than less phenomenologically desirable theories in $(2,1)$ or
$(5,1)$. This universe ${\cal I}$ may also have some interesting
cosmological features: for example, the 3-brane and 7-brane
cosmology of  \cite{Karch:2005yz}.

Finally, we recall \cite{Duff2:2006} that although the flipping of
the sign of the metric tensor may leave the equations of motion
invariant, a choice has to be made when choosing the boundary
conditions. A metric vacuum expectation value
\be
\langle G_{MN}(x) \rangle = \eta_{MN}
\ee
breaks the reversal symmetry spontaneously. Similarly the dilaton
field expectation value
\be
\langle {\Hdila}(x) \rangle = g_{s}
\ee
breaks spontaneously the sign reversal ${\Hdila} \rightarrow
-{\Hdila}$. So one might entertain the possibility of different 
domains within a given universe.


\end{document}